\documentclass[runningheads]{llncs}
\usepackage{graphicx}
\usepackage{amsmath,amssymb}
\usepackage{hyperref}

\title{Reconsidering Requirements Engineering: Human--AI Collaboration in AI-Native Software Development}
\titlerunning{Human-AI Collaboration in Requirements Engineering}

\author{Mateen Ahmed Abbasi\inst{1} \and
Petri Ihantola\inst{1} \and
Tommi Mikkonen\inst{1} \and
Niko Mäkitalo\inst{1}}

\authorrunning{M. A. Abbasi et al.}

\institute{Faculty of Information Technology, University of Jyväskylä, Jyväskylä, Finland\\
\email{\{mateen.a.abbasi, petri.j.ihantola, tommi.j.mikkonen, niko.k.makitalo\}@jyu.fi}}

\begin{document}

\maketitle

\begin{abstract}
Requirement Engineering (RE) is the foundation of successful software development. In RE, the goal is to ensure that implemented systems satisfy stakeholder needs through rigorous requirements elicitation, validation, and evaluation processes. Despite its critical role, RE continues to face persistent challenges, such as ambiguity, conflicting stakeholder needs, and the complexity of managing evolving requirements. A common view is that Artificial Intelligence (AI) has the potential to streamline the RE process, resulting in improved efficiency, accuracy, and management actions. However, using AI also introduces new concerns, such as ethical issues, biases, and lack of transparency.  
 This paper explores how AI can enhance traditional RE practices by automating labor-intensive tasks, supporting requirement prioritization, and facilitating collaboration between stakeholders and AI systems. The paper also describes the opportunities and challenges that AI brings to RE. In particular, the vision calls for ethical practices in AI, along with a much-enhanced collaboration between academia and industry professionals. The focus should be on creating not only powerful but also trustworthy and practical AI solutions ready to adapt to the fast-paced world of software development.

\end{abstract}

\section{Introduction}

Requirements engineering (RE) is the foundation of software development. During RE, stakeholders' needs are transformed into explicit 
system requirements \cite{pohl2010}. The success of software project 
depends on 
the quality of the RE process \cite{konrad2008requirements}. This critical initial phase involves eliciting, analyzing, specifying, and validating the needs, constraints, and expectations of stakeholders to establish a clear understanding of the software system 
\cite{umber2012requirements,konrad2008requirements}. It also requires understanding the problem context, modeling requirements, and reconciling conflicting demands\cite{sommerville2011software}. However, RE faces persistent challenges, including ambiguity in requirements\cite{bano2015addressing}, conflicts among stakeholders, and the dynamic nature of software development\cite{dasanayake2019impact}. Effective RE significantly reduces rework and improves project success \cite{hall2002requirements}.

AI is emerging as a transformative force in software engineering, particularly in RE\cite{Dalpiaz2020}, offering new opportunities to support or automate RE activities such as requirement elicitation, traceability, and task prioritization \cite{hudaib2018requirements,guo2024natural,bosch2021engineering}. However, the integration of AI into RE also brings a new set of risks. The complexity and unpredictability of AI systems can introduce uncertainty into software projects \cite{bosch2021engineering}, particularly when RE depends heavily on human understanding, context, and negotiation. Interpreting AI-generated recommendations becomes critical for aligning system behavior with stakeholder expectations. 

In this paper, we focus on data-driven AI techniques, particularly Machine Learning (ML) and Generative AI approaches, and explore how they are shaping the future of RE. While these methods offer significant potential, they also raise challenges related to data quality, interpretability \cite{Felderer2021}, and ethical concerns such as inherited bias in AI-generated outputs \cite{Yarger2020,Ferrara2023,Nazer2023,Hajian2016}. These shifts suggest that AI is not just improving RE tasks in isolation, but gradually transforming the way RE is practiced, how roles are defined, and how decisions are made throughout the development lifecycle.

This article contributes to this discussion by:
\begin{itemize}
    \item Identifying long-standing RE challenges,
    \item Analyzing how AI is affecting and reshaping these challenges,
    \item Outlining future research directions aimed at mitigating AI-driven challenges in RE.
\end{itemize}

The rest of the paper is structured as follows. Section~\ref{sec:Methodology} describes the methodology.  Section~\ref{sec:challenges} discusses key RE challenges, the role of AI in addressing them, and new challenges introduced by AI. 
Section~\ref{sec:discussion} explores potential solutions and directions for future research. 
Section~\ref{sec:conclusions} presents our conclusions.

\section{Methodology}
\label{sec:Methodology}


To examine how AI technologies are affecting the requirements engineering process, we conducted a structured (though not fully systematic) literature review designed to be transparent and reproducible. Drawing from established literature review practices, the process applied defined search strategies, relevance-based inclusion criteria, and thematic synthesis to ensure consistency and traceability across all phases. The research process followed three phases (I, II, and III) as explained below.

\subsection{Phase I: Identifying Traditional RE Challenges}



In Phase I, we conducted a literature review to identify the key challenges associated with traditional requirements engineering. Our search focused on databases including Google Scholar and Scopus. The search strings included terms such as "requirements engineering challenges", "common issues in RE", and "RE problem studies". We prioritize peer-reviewed publications and filtered out those that did not directly address RE challenges. Studies were screened according to the relevance of the title and abstract, followed by full text evaluation. 

To group RE challenges into the five key themes discussed in Section~\ref{sec:challenges}, we used qualitative coding and thematic grouping. Each selected study was fully reviewed and relevant challenges were tagged and grouped based on recurrence and thematic similarity. For example, challenges related to vague requirements, inconsistent terminology, and stakeholder misunderstandings were grouped under "Ambiguity and Conflicting Requirements." The search was concluded upon reaching thematic saturation, meaning that additional articles no longer introduced new or substantially different categories of challenges. In total, approximately 35 peer-reviewed articles were reviewed in full for thematic synthesis.

Numerous challenges in traditional RE have been identified in previous studies, including difficulties in requirements elicitation \cite{de2012elicitation,umber2012requirements}, ambiguous and conflicting requirements \cite{bano2015addressing}, communication barriers \cite{schmid2014challenges,de2012elicitation}, dynamic and volatile requirements \cite{biddle20061,dasanayake2019impact,nurmuliani2004analysis}, poor traceability \cite{cleland2012software,antoniol2017grand,guo2024natural}, and issues with prioritization and stakeholder involvement \cite{abdelazim2020framework,hudaib2018requirements}. Other important challenges include insufficient domain knowledge \cite{birk2007challenges}, feedback cycles in requirement validation \cite{bano2013service}, requirement reuse in inappropriate contexts, inadequate modeling of non-functional requirements \cite{shah2014review}, and evolving requirements during development \cite{kasauli2021requirements}.

The RE challenges discussed here were selected based on their prevalence, significance, and impact across RE practices.
Our selection process involved:

\begin{itemize}
\item Reviewing peer-acknowledged literature frequently included in systematic reviews and key RE studies.
\item Prioritizing research that explicitly analyzed major RE challenges.
\item Ensuring coverage of challenges that remain critical across various RE methodologies (waterfall, agile, hybrid models).
\end{itemize}

After identifying the challenges from the following literature reviews \cite{alsalemi2017systematic,bano2015addressing,schmid2014challenges,shahbeklu2024requirement,tukur2021requirement} and empirical studies \cite{alqaisi2018effects,bano2015addressing,hall2002requirements,damian2003insight,shahbeklu2024requirement} we synthesized the findings to form a comprehensive and balanced understanding of traditional RE challenges.
The results have been introduced in Section~\ref{sec:challenges}.

\subsection{Phase II: Review Process on AI in RE}
In Phase II, we examined recent research exploring how AI techniques are applied within requirements engineering. 
We followed the following review process:

\begin{itemize}
\item Identification: 
To investigate how AI is applied to address RE challenges, we searched Google Scholar using the following query: "AI in Requirements Engineering" OR "Natural Language Processing for RE" OR "Machine Learning in RE" OR "AI for Requirement Prioritization" OR "Automated Requirement Traceability." This initial search yielded 136 results. Although Phase I used both Scopus and Google Scholar, Phase II relied solely on Google Scholar due to its broader coverage of recent and preprint literature, which is especially valuable in the rapidly evolving field of AI. We acknowledge that using a single source may introduce selection bias due to less curation compared to databases like Scopus. To mitigate this, we applied an iterative refinement strategy that included removing redundant entries, analyzing backward and forward citations from key articles, adjusting search keywords, and prioritizing the most recent peer-reviewed publications directly related to AI in RE. Title and abstract screening were used to exclude studies focusing solely on theoretical discussions or on non-functional requirements. We reviewed the full text of the remaining papers to ensure methodological relevance. This filtering process resulted in a curated set of studies that formed the basis for the analysis in Phase III.

\item Eligibility: We focused on studies that proposed, evaluated, or discussed AI techniques such as Natural Language Processing (NLP), Machine Learning (ML), Deep Learning (DL), and Information Retrieval methods applied to RE tasks including requirement elicitation, ambiguity detection,  traceability automation, conflict resolution, prioritization techniques, and ethical concerns such as bias mitigation and explainability. 
\item Inclusion: We selected studies that directly focused on the integration of AI techniques with RE for detailed analysis. These selected studies were supplemented by seminal papers frequently cited in RE research to ensure a comprehensive understanding of existing challenges, advancements, and gaps in AI-driven RE methodologies. These findings set the foundation for our later exploration of emerging AI approaches in Section~\ref{sec:discussion}.
\end{itemize}
\subsection{Phase III: Analysis How AI is Changing RE Challenges}
In Phase III, we analyzed how AI techniques can help address specific RE challenges, based on insights from the literature. The evaluation criteria included:
\begin{itemize}
    \item Accuracy and Reliability: How effectively AI models can identify, classify, and refine requirements.
    \item Scalability: The capacity of AI tools to manage large and evolving sets of requirements.
    \item Transparency and Explainability: The extent to which AI decisions can be interpreted by RE professionals.
    \item Ethical Considerations: How well the models mitigate bias and promote fairness.
\end{itemize}
While our review is not exhaustive, we focused on synthesizing the most influential and recent works from both theoretical and empirical perspectives to provide a balanced discussion on AI integration in RE processes. 

From Phase I, we identified a broad range of traditional RE challenges. For structured analysis, we narrowed our focus to five recurring and thematically central challenges: (1) ambiguity and conflicting requirements, (2) dynamic and volatile requirements, (3) communication barriers, (4) poor traceability, and (5) prioritization and stakeholder involvement. These challenges were selected based on the frequency with which they appear in the literature and their relevance to AI-based solutions. In this phase, we examined how AI addresses these challenges and also identified new issues that AI introduces. Section~\ref{sec:challenges} is structured accordingly.

While some concerns such as data bias or explainability are common in general AI applications, they are particularly impactful in RE due to the foundational role of requirements in shaping the entire software development lifecycle. Misinterpretations at this stage can propagate into critical downstream flaws.

\section{RE Challenges and the Impact of AI}
\label{sec:challenges}
This section presents a unified analysis of key challenges in RE and examines how Artificial Intelligence (AI) techniques are influencing these areas. For each challenge, we explore the traditional difficulties faced in RE, how AI technologies have been used to address them, the new issues introduced through AI integration, and practical examples where applicable. This synthesis draws on findings from both classical RE literature and recent research on AI applications in RE.

\subsection{Ambiguity and Conflicting Requirements}
Ambiguity and conflicting requirements have been one of the major challenges in RE, due to vague stakeholder inputs and different interpretations \cite{bano2015addressing,Dalpiaz2020}. Ambiguity and conflicting requirements arise when stakeholders provide ambiguous, incomplete, or inconsistent information \cite{shah2015resolving,sandhu2015state}, These problems generally occur due to differences in understanding and unclear requirements among stakeholders \cite{dube2010process,sandhu2015state}. Different stakeholders may use different terminology or may have conflicting objectives, which makes requirements vague or inconsistent \cite{nuseibeh2000requirements,maiden2004provoking,tukur2021requirement}. Ambiguous requirements increase the probability of errors in later development phases \cite{shah2014review}. Stakeholders are usually unable to clearly articulate all their requirements during the early stages of a project, leading to incomplete requirements. As development proceeds, the missing requirements are realized during the system integration or testing phases \cite{firesmith2007common}.

\textbf{\textit{How AI has changed this challenge:}}
 AI-based methods can handle large volumes of requirements and ensure consistency between artifacts \cite{Dalpiaz2020}. LLMs act as conversational agents that can process domain-specific knowledge and generate structured requirements from unstructured data\cite{arora2024advancing}. NLP-based tools such as CoreNLP, NLTK, and OpenNLP assist in detecting ambiguous phrases and inconsistencies in requirements\cite{alzayed2021bird}. However, resolving these ambiguities requires human expertise to ensure correctness and alignment with stakeholder expectations \cite{alzayed2021bird,ferrari2019nlp}. AI tools can assist in extracting requirements from various sources, such as emails, meeting notes, and documentation. This helps to resolve the issues of incomplete requirements, ensuring that no important details are missed \cite{arora2024advancing}.

\textbf{\textit{What new challenges AI has introduced:}}
While AI reduces ambiguity, it also introduces problems such as biases in language models. One key challenge is the need for large and high-quality datasets for training and evaluating AI models. The effectiveness of AI models in RE depends on the quality and diversity of the data used for training \cite{ghaisas2024dealing}. Additionally, prompt design plays a critical role in requirement extraction using LLMs since poorly constructed prompts may cause misinterpretation or amplify biases \cite{arora2024advancing}. Collecting and preparing proper datasets for AI systems is resource-intensive and requires careful curation to ensure relevance and accuracy. If the data used to train the models is biased or incomplete, the AI might misunderstand stakeholders' requirements or fail to detect conflicts \cite{Ferrara2023}. 

\textbf{\textit{Example:}} AI-driven requirement prioritization can struggle when it comes to managing conflicting stakeholder needs, especially when requirements are ambiguous. 
For instance, an AI system focused on optimizing functional performance may deprioritize usability requirements, assuming they are less critical. However, stakeholders could have different opinions on the relative priority of performance versus user experience, leading to overlooked priorities and mismatched expectations.

\subsection {Dynamic and Volatile Requirements}
The ever-changing nature of requirements, often referred to as “requirements volatility,” poses a significant challenge in software development\cite{biddle20061}. Volatile requirements can disrupt project timelines, increase costs, lead to significant rework, and affect stability and consistency in the development process \cite{biddle20061,dasanayake2019impact}. Traditional RE struggles with managing frequent requirement changes due to evolving business needs. Excessive change may also affect system design, project scope, and budget \cite{nurmuliani2004analysis,galster2017variability}. Another challenge arises due to stakeholders' shifting needs, which inevitably bring changes in requirements during the development process\cite{almeida2017challenges}. Dynamic and frequent changing requirements can derail the success of the project \cite{tukur2021requirement}.

\textbf{\textit{How AI has changed this challenge:}}
ML algorithms can address the challenges of dynamic and volatile requirements through predictive analytics \cite{alsalemi2017systematic}, which anticipate possible requirement changes based on past projects and stakeholder behavior \cite{iqbal2018bird,Dalpiaz2020}. AI solutions can help reduce the effects of requirement changes by offering better tools in areas like requirements elicitation, analysis, and change management \cite{Dalpiaz2020}.

\textbf{\textit{What new challenges AI has introduced:}}
AI may offer promising solutions, it also brings its own set of challenges \cite{gjorgjevikj2023requirements}. AI models trained on historical data can assist in identifying the most important requirements by predicting how they will impact cost, time, and user satisfaction \cite{harman2012search}. Predictive analytics relies on the quality of existing data. If the data is incomplete or outdated, the predictions might be incorrect or misleading \cite{gjorgjevikj2023requirements}. Over-reliance on AI predictions might cause teams to overlook new factors that may not align with historical trends \cite{aldoseri2023re}. One significant challenge is that volatility, often driven by changing contexts or evolving user needs, cannot be entirely predicted or controlled through AI. Another limitation is the inherent complexity of AI models. AI models trained on biased datasets, tend to perpetuate the already existing biases in requirements elicitation and prioritization. The biggest concern is how fair an AI-driven RE process will be, especially in ensuring equitable treatment of stakeholders, while minimizing bias in interpretations \cite{mehrabi2021survey}. AI systems should be held accountable for ethical norms governing biases and transparency \cite{boch2022}.

\textbf{\textit{Example:}} A development team working on an e-commerce platform uses predictive analytics to identify future needs in a fast-paced business area. However, as the user base of the platform evolves (e.g. demographics of the region), the training data no longer reflects the changing realities of the platform. The predictions become less accurate and start overemphasizing some historical trends while missing emerging ones (e.g., new payment methods or region-specific regulations).

\subsection {Communication Barriers}
Communication barriers arise when stakeholders and development teams face problems in sharing and understanding information \cite{connor2009bridging}, particularly during the elicitation of requirements, where misunderstandings and incomplete information are common.  Differences in technical knowledge, language, and culture hinder communication between stakeholders and developers \cite{de2012elicitation}. 
These differences can lead to incomplete or misinterpreted requirements, ultimately affecting project outcomes. 
Global teams often encounter cultural and language barriers that complicate collaboration\cite{schmid2014challenges,damian2003insight}.

\textbf{\textit{How AI has changed this challenge:}}
AI can enhance communication and collaboration in the requirements engineering process \cite{alzayed2021bird}. One of the primary way to reduce communication barriers is through language translation and NLP techniques \cite{alzayed2021bird,chen2024conversational}. AI-driven chatbots and virtual assistants significantly improve communications by providing standard responses to common queries and helping clarify requirements during discussions \cite{chen2024conversational}.

\textbf{\textit{What new challenges AI has introduced:}} AI brings significant benefits in addressing communication barriers in RE, it also introduces several challenges that must be considered \cite{gjorgjevikj2023requirements,Dalpiaz2020}. One major issue is bias within AI systems \cite{schwartz2021proposal}. AI systems may inherit biases from training data and can skew the interpretations of requirements\cite{mehrabi2021survey}. AI solutions for communication barriers can lead to over-reliance on automated translations, which may not be able to convey cultural nuances or context-specific meanings \cite{kirkpatrick2020natural}. Moreover, misinterpretations by AI tools can even exacerbate rather than mitigate communication issues in high-stakes discussions\cite{nijiati2020problems}.

\textbf{\textit{Example:}} AI chatbots and NLP tools used to bridge communication gaps between technical and non-technical stakeholders can sometimes misunderstand special terms or cultural details. For example, in a global project, one stakeholder might say a "warehouse" means a main storage place, while another might mean a decentralized distribution hub. If the AI fails to recognize this context, it may generate specifications that don't align with expectations, exacerbating communication issues rather than solving them.

\subsection {Poor Traceability}
Requirement traceability is important to establish the connections between requirements, design, and implementation\cite{antoniol2017grand}. Poor traceability refers to the difficulty in establishing and maintaining the link between requirements and other development artifacts, such as design documents, code, and test cases\cite{guo2024natural,antoniol2017grand}. Manual traceability is error-prone and labor-intensive, resulting in gaps in requirements management \cite{cleland2012software}. The lack of traceability complicates validation and verification, making it harder to ensure that the final product satisfies all requirements. 

\textbf{\textit{How AI has changed this challenge:}}
Deep learning and NLP techniques can be used to automate traceability linking between requirements and other artifacts\cite{guo2017semantically}. The integration of AI techniques into the RE process can provide benefits by enhancing traceability and reducing manual effort \cite{cleland2012software}.

\textbf{\textit{What new challenges AI has introduced:}}
The quality and reliability of AI-generated traceability links are highly dependent on the quality of the training data. If the data is incomplete or biased, it may result in recommendations that would harm the overall traceability of the system \cite{gjorgjevikj2023requirements}. Errors in AI-generated traceability links might go unnoticed if there is not enough human oversight, which could cause problems later in the development process \cite{guo2017semantically}. Another key challenge is ensuring the traceability of AI itself, such as tracking the sources of training data and models, to promote transparency and accountability\cite{luthi2020distributed}.

\textbf{\textit{Example:}} If the AI automatically links requirements to code and test cases but incorrectly maps them, it could create false dependencies. For example, it might associate a requirement for "data encryption" with unrelated logging functions, leading to wasted effort in verifying irrelevant links. Also, relying too much on the AI’s traceability suggestions might lower manual checks, increasing the likelihood of undetected errors.

\subsection {Prioritization and Stakeholder Involvement} 
Prioritization ensures that the most critical features are addressed early in the software development process, prioritization and stakeholder involvement are critical to ensure essential requirements align with the project objectives to be delivered first\cite{abdelazim2020framework,hudaib2018requirements}. It becomes challenging when stakeholders have conflicting interests, resources are limited, or some are not actively involved in the RE process\cite{alqaisi2018effects,hudaib2018requirements}. Requirement prioritization is inherently subjective, as stakeholder interests often vary. Limited participation of stakeholders in RE phases may result in requirements that do not meet user expectations \cite{hudaib2018requirements}.

\textbf{\textit{How AI has changed this challenge:}}
The identification and involvement of the right stakeholders is crucial \cite{hudaib2018requirements}. NLP  and ML techniques can help by automating document analysis, user feedback, and other sources to identify key stakeholders and their needs \cite{felfernig2021ai,maalej2023tailoring}. AI techniques like utility-based recommendation, matrix factorization, and content-based recommendations can analyze requirements and support prioritization \cite{felfernig2021ai}. These AI techniques can automate the prioritization process, stakeholder preferences, requirement dependencies, and changing priorities over time \cite{felfernig2021ai,hudaib2018requirements}. 
Techniques such as utility-based recommendation and optimization approaches like Analytic Hierarchy Process (AHP) and Binary Search Tree \cite{hudaib2018requirements,felfernig2021ai}  can be enhanced through modern ML algorithms such as Random Forests and Linear Programming for analyzing trade-offs between cost, time, and stakeholder needs to achieve optimal prioritization.

\textbf{\textit{What new challenges AI has introduced:}}
AI tools and techniques can enhance prioritization and stakeholder involvement, but it also raises issues such as bias, transparency, and ethical considerations \cite{Ferrara2023,santhanam2020quality,ghaisas2024dealing,belani2019requirements}. As ML-based AI systems rely on historical data to improve prioritization, Such data often fail to capture the specific context of the project\cite{santhanam2020quality}. Moreover, AI-driven tools mostly work as black boxes \cite{handelman2019peering}, making it difficult for users to understand how decisions are made. This lack of transparency can lead to low trust in AI recommendations\cite{bach2024systematic,doshi2017towards}. The use of AI in RE also raises ethical concerns related to accountability and decision-making. Ethical accountability is essential because AI systems might reinforce biases, unfairly prioritize requirements, or make choices that conflict with the real needs of the stakeholders \cite{Ferrara2023}. Accountability in AI can improve trust and collaboration between stakeholders by clarifying decision processes \cite{ribeiro2016should}, especially in large projects with many stakeholders \cite{caliskan2017semantics}. It has to be ensured that AI recommendations comply with ethical standards and stakeholder values\cite{floridi2018ai4people}.

\textbf{\textit{Example:}} AI tools that analyze and prioritize requirements might unintentionally favor stakeholders who provide more data or have more consistent feedback patterns. For example, in a project involving multiple teams, AI might prioritize features requested by the marketing department due to clearer documentation while neglecting crucial but less well-documented feedback from other stakeholders. This bias could result in misaligned prioritization that does not adequately represent needs of all stakeholders. Table 1 summarizes the key RE challenges, the ways in which AI addresses them, and the new challenges that AI introduces.

\begin{table*}[ht]
\small
\centering
\caption{Summary of Challenges and AI Impact}
\begin{tabular}{|p{3.8cm}|p{4.3cm}|p{4.3cm}|}
\hline
\textbf{RE Challenge} & \textbf{How AI Addresses It} & \textbf{New Challenges Introduced by AI} \\
\hline
Ambiguity and Conflicting Requirements & NLP tools and LLMs structure unstructured input and detect ambiguity. & Bias from training data, reliance on prompt engineering, and limited explainability. \\
\hline
Volatile and Dynamic Requirements & Predictive analytics models forecast requirement changes and adjust priorities. & Overfitting to historical trends, difficulty capturing emerging needs. \\
\hline
Communication Barriers & Chatbots and translators ease understanding across diverse teams. & Loss of context or nuance in translation, misinterpretation of domain-specific terms. \\
\hline
Poor Traceability & Automates linking between requirements, code, and tests. & Errors in automated linking, lack of transparency in traceability logic. \\
\hline
Prioritization and Stakeholder Involvement & ML-based ranking systems analyze feedback and optimize prioritization. & Bias favoring well-documented or dominant stakeholders, lack of transparency in decision-making. \\
\hline
\end{tabular}
\label{tab:ai_re_challenges_column}
\end{table*}

\section{Discussion and Future Directions}
\label{sec:discussion}


 

This paper set out to examine how Artificial Intelligence is influencing Requirements Engineering at a broader level, not only by offering technical solutions, but by gradually reshaping the way RE is performed, who performs it, and how decisions are made throughout the process. 

\begin{figure}[ht]
\centering
\includegraphics[width=0.48\textwidth]{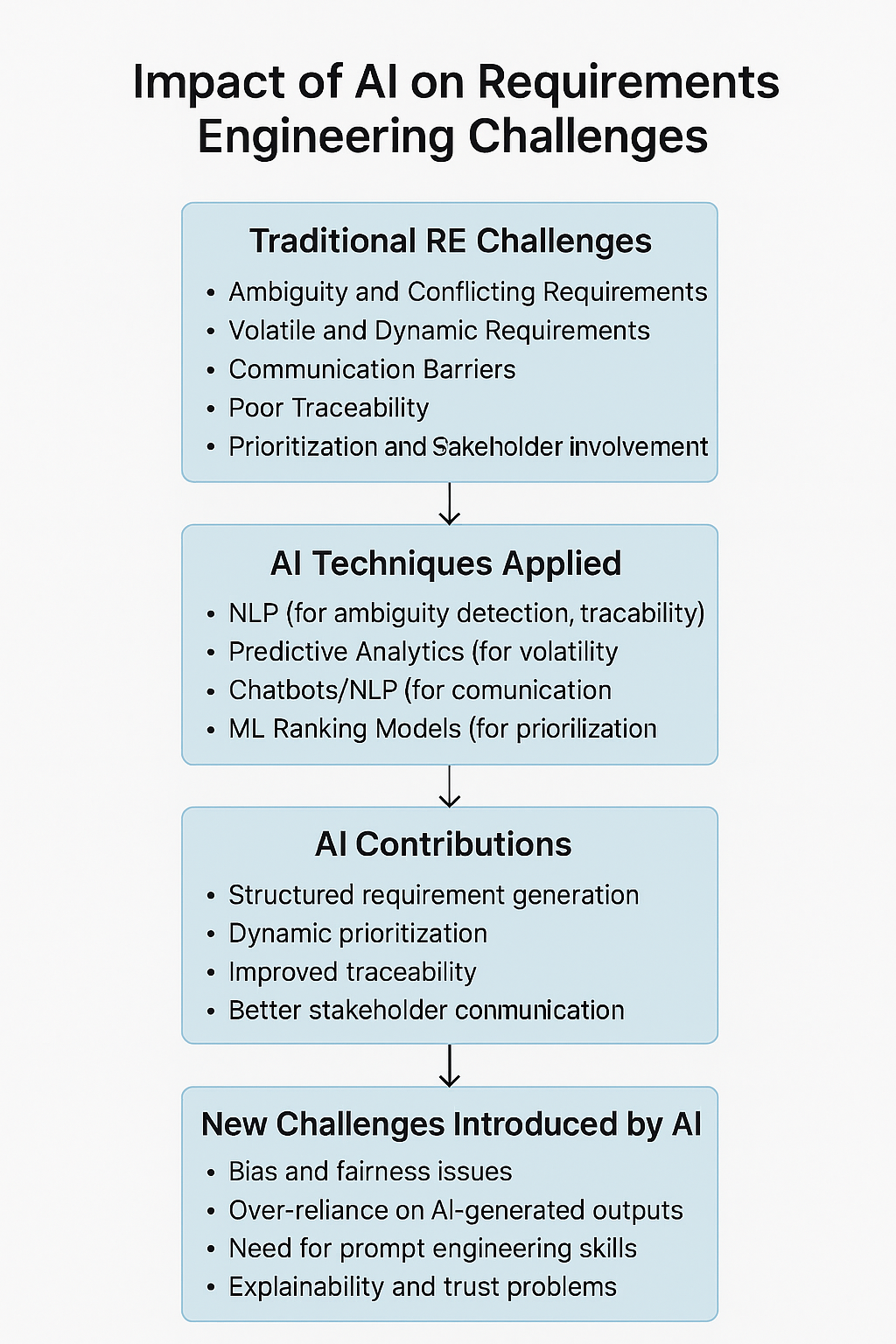}
\caption{Conceptual flow linking RE challenges, AI contributions, and new challenges introduced by AI integration.}
\label{fig:conceptual_flow}
\end{figure}

Our analysis shows that AI is no longer just assisting with individual RE tasks; it's starting to affect core aspects of the RE process, including ambiguity resolution, stakeholder prioritization, traceability, and adaptation to change. These shifts also raise important concerns about the role of human oversight, explainability of AI decisions, and the trust that stakeholders place in automated recommendations. Figure~\ref{fig:conceptual_flow} illustrates how traditional RE challenges intersect with AI-driven approaches, and how new challenges emerge as AI becomes more embedded in RE practices.

While issues such as dataset adequacy, over-reliance on AI-generated outputs, prompt engineering skills, bias and fairness concerns, and explainability and trust problems are broadly recognized in AI research, they manifest with particular importance in the context of Requirements Engineering. Unlike many other domains, RE heavily relies on nuanced, context-rich human inputs, where misinterpretations, biases, or lack of transparency during early stages can propagate into critical system specifications and significantly impact downstream development activities. 

Moreover, our literature review revealed notable gaps: although various AI methods have been applied to traditional RE tasks, there remains limited exploration of emerging approaches such as retrieval-augmented generation (RAG), lightweight domain adaptation techniques, and dynamic prioritization using reinforcement learning.  In this section, we discuss how AI techniques can help mitigate these challenges, and present our future research ideas and objectives.

\subsection{AI Techniques for Enhancing RE}

\textbf{\textit{Retrieval-Augmented Generation (RAG):}} Retrieval-Augmented Generation (RAG) combines retrieval techniques with generative models to produce contextually relevant outputs \cite{gupta2024comprehensive}. In RE, RAG can be utilized to retrieve knowledge from requirement specifications, past project documentation, and knowledge bases of specified domains. 

\begin{itemize}
    \item \textit{Addressing RE Challenges:} RAG can help mitigating ambiguous and conflicting requirements by retrieving structured requirement statements from past projects, which reduces inconsistencies and improves clarity.

    \item \textit{Future Research:} We are developing a RAG-based tool to help software development teams extract past requirements from project repositories in a systematic and reusable way for multiple projects. Then evaluate how RAG can help to resolve these issues. 
\end{itemize}


\textbf{\textit{Fine-Tuning Pre-trained Models:}} Fine-tuning involves modifying pre-trained language models to perform more effectively in particular domains by training them with domain-relevant datasets\cite{liu2021autofreeze}. Fine-tuning large language models (LLMs) like BERT and GPT for RE activities can improve the accuracy in eliciting, classifying, and analyzing requirements. This can better handle domain-specific terminologies and stakeholder preferences by training on RE datasets. 

Though fine-tuning can enhance model performance for RE tasks, it also presents challenges such as high computational costs, the risk of overfitting, and environmental concerns. Therefore, our future work will also explore more lightweight domain adaptation strategies, such as prompt optimization, adapter-based fine-tuning, and retrieval-augmented approaches, to balance domain specificity with practical deployment feasibility.

\begin{itemize}
    \item \textit{Addressing RE Challenges:} Fine-tuned models can improve requirement elicitation and traceability by generating more accurate and structured requirement documents.

    \item \textit{Future Research:} We will explore different fine-tuning approaches and study how fine-tuned model can help tackling regulatory compliance challenges (e.g., HIPAA, GDPR) so that AI-driven requirements assistant produce requirements that are regulation compliant.
\end{itemize}




\textbf{\textit{Reinforcement Learning (RL):}} 
In an software development project, a reinforcement learning (RL) system can continuously learn from feedback over sprints (rewards or penalties), and refine priority suggestions for requirements to align with changing user needs. 

\begin{itemize}
    \item \textit{Addressing RE Challenges:} Reinforcement learning methods can help improve decision-making in RE by managing dynamic and volatile requirements and by learning from past requirements changes and predicting future changes to enable more adaptive and effective requirement management.

    \item \textit{Future Research:} In future, we study how RL methods can be used to help with requirement prioritization by learning from feedback experience on previous projects. By defining reward functions in terms of cost, risk, and stakeholder satisfaction, RL agents can assist in changing requirement priorities as projects evolve\cite{bagherzadeh2021reinforcement}.
\end{itemize}

\subsection{Research Objectives}

\textbf{\textit{Empirical Testing of AI Models in RE Tasks:}} 
Our objective is to evaluate how efficient AI techniques such as RAG, fine-tuning, and RL can help mitigate the identified RE challenges. 
In future studies, we plan to conduct 
benchmarking the AI methods with traditional RE processes to measure improvements in efficiency, accuracy, and stakeholder satisfaction.

\textbf{\textit{Comparative Analysis of AI-driven vs. Traditional RE Processes:}} Our objective is to compare AI-assisted and manual RE processes to analyze where AI is most valuable and where human decision-making is still very pertinent. We plan to use surveys, controlled experiments, and performance analysis to compare the outcomes of each approach.

\textbf{\textit{Developing a Concrete Integration Framework:}} Our final objective is 
the creation of a comprehensive framework to integrate AI into industrial RE practices. The framework will outline best practices for handling data, improving models, interpreting results, and incorporating human input into the process. It also needs to provide guidelines on how to balance automation with ethics and achieve transparency and accountability in AI-supported RE activities. 

\subsection{Implications for Practice and Theory}

\textbf{\textit{Implications for Practice:}} The findings of this article show how AI can be applied to real RE problems, such as using NLP tools to detect ambiguous requirements, ML models for prioritizing tasks, and reinforcement learning to adapt to changing needs. These techniques can help reduce manual work, improve accuracy, and make the RE process more manageable in practice. The discussion of current tools and emerging strategies can help guide the selection or development of AI-based RE solutions in real-world projects.

\textbf{\textit{Implications for Theory:}} This work contributes to the ongoing research connecting AI with RE by pulling together current studies and identifying gaps, like explainability, ethical concerns, and data quality in AI-based RE tools. It also sets the stage for future studies that will experiment with how effective AI is in RE, especially through novel methods like RAG and reinforcement learning.



\section{Conclusions}
\label{sec:conclusions}


This paper presents a structured analysis of the evolving role of AI in RE, highlighting both its transformative potential and the new challenges it brings. Techniques like RAG, fine-tuning, and reinforcement learning offer promising approaches to address key RE challenges, including ambiguity, volatility, and prioritization. Integrating AI into RE processes can help organizations reduce time and costs while improving accuracy and reliability. However, despite these benefits, numerous issues like ethical concerns, algorithmic biases, transparency, and fairness arise. 

Our findings suggest that AI should be considered as an augmentative tool, not a substitute for human decision-making in RE. AI can enhance efficiency, but its results should be interpretable, fair, and aligned with stakeholder needs. To address the new challenges AI introduces into RE, our future work will focus on exploring techniques to detect and reduce biases in AI-generated requirements, ensuring fairness in requirements prioritization and decision-making.

\begin{credits}
\subsubsection{\ackname} This work has been supported by FAST, the Finnish Software Engineering Doctoral Research Network, funded by the Ministry of Education and Culture, Finland.
\subsubsection{\discintname}
The authors have no competing interests to declare that are relevant to the content of this article.
\end{credits}

\bibliographystyle{IEEEtran}
\bibliography{References}

\end{document}